\documentclass[aps,prd,reprint,groupedaddress,showpacs,nofootinbib,amsmath]{revtex4-1}

\usepackage{amsmath}
\usepackage{amssymb}
\usepackage{bbold}
\usepackage{graphicx}
\usepackage{verbatim}
\usepackage{hyperref}
\usepackage{indentfirst}
\usepackage{mathtools}
\usepackage{xcolor}
\usepackage{color}
\usepackage{adjustbox}
\usepackage{placeins}
\usepackage{lipsum}
\usepackage{csquotes}

\usepackage[font=small,labelfont=bf,labelsep=space]{caption}
\begin{document}

\title{Cosmological models with asymmetric quantum bounces}

\author{P.~C.~M.~Delgado}\email{pmordelgado@gmail.com}
\affiliation{CBPF - Centro Brasileiro de
Pesquisas F\'{\i}sicas, Xavier Sigaud st. 150,
zip 22290-180, Rio de Janeiro, Brazil.}

\author{N.~Pinto-Neto}\email{nelson.pinto@pq.cnpq.br}
\affiliation{CBPF - Centro Brasileiro de
	Pesquisas F\'{\i}sicas, Xavier Sigaud st. 150,
	zip 22290-180, Rio de Janeiro, Brazil.}

\date{\today}

\begin{abstract}
In quantum cosmology, one has to select a specific wave function solution of the quantum state equations under consideration in order to obtain concrete results. The simplest choices have been already explored, in different frameworks, yielding, in many cases, quantum bounces. As there is no consensually established boundary condition proposal in quantum cosmology, we investigate the consequences of enlarging known sets of initial wave functions of the universe, in the specific framework of the Wheeler-DeWitt equation interpreted along the lines of the de Broglie-Bohm quantum theory, on the possible quantum bounce solutions which emerge from them. In particular, we show that many asymmetric quantum bounces are obtained, which may incorporate non-trivial back-reaction mechanisms, as quantum particle production around the bounce, in the quantum background itself. In particular, the old hypothesis that our expanding universe might have arisen from quantum fluctuations of a fundamental quantum flat space-time is recovered, within a different and yet unexplored perspective.
\end{abstract}

\maketitle

\tableofcontents

\section{Introduction}
According to the Penrose-Hawking singularity theorems in General Relativity \cite{penrose-hawking}, the universe has a beginning described by a singularity in
space-time, which is outside the scope of the theory and, hence, cannot be investigated. This led
to the idea that, in this extreme domain, characterized by very high energy densities and curvature, General Relativity
must undergo modifications, which may be due to quantum gravitational effects. Therefore, it is
necessary to formulate a quantum theory of gravity to describe the domain previously held as a singularity. 

Quantum Mechanics, on the other hand, is understood as a fundamental theory able to describe any
physical system, including the whole universe. However, the Copenhagen interpretation cannot be
applied to cosmology. The reason is that, in order to solve the measurement problem, this interpretation postulates that the wave function collapses when an observer performs a measurement on the system. Thus an external classical domain is required to perform the collapse of the
wave function. 

There are some proposals to circumvent this conceptual problem, the most famous being the Many-Worlds interpretation \cite{many-worlds}, the spontaneous collapse approach \cite{spontaneous-collapse}, and the de Broglie-Bohm quantum theory \cite{Bohm:1951xw,Bohm:1951xx}. We will adopt this last one, a deterministic interpretation in
which real trajectories in the configuration space exist. The probabilistic character of Quantum Mechanics is due to the existence of hidden variables (initial field configurations), and arises statistically.
In this theory, 
the collapse of the wave function is effective: 
the system occupies one of the branches of the wave function, and the others remain empty and incommunicable to each other. 
Therefore, an external observer is no longer needed, and we achieve the conceptual coherence necessary to apply this approach to cosmology.

The quantum cosmological models that arise from this approach enable the avoidance of the initial singularity,
giving rise to a bounce \cite{Pinto-Neto:2013toa,PintoNeto:2004uf}, or even multiple bounces \cite{Peter:2016kan,Bacalhau:2017qnu}, which are preceded by a contraction of the scale factor and followed by an expanding phase. 

In this paper, we consider generalizations of the quantum cosmological models found in Refs~\cite{Pinto-Neto:2013toa,PintoNeto:2004uf} arising from the Wheeler-DeWitt quantization of the background, which are symmetric around the bounce, obtained from enlarged prescriptions for the initial wave function. Our aim is to obtain asymmetric bounces, capable to describe non-linear back-reactions coming from particle production around the bounce, which can alter the background evolution in the expanding phase. Indeed, taking into account generalizations of the initial Gaussian wave functions considered in Refs~\cite{Pinto-Neto:2013toa,PintoNeto:2004uf}, we were able to obtain a variety of asymmetric quantum bounce trajectories in different contexts, with quite interesting properties, as it will be discussed in the sequel. 

The paper is divided as follows: in the next section we present the mini-superspace model in which the de Broglie-Bohm quantization will be implemented, and the standard symmetric quantum bouncing trajectories obtained from initial Gaussian wave functions centered at the origin, and without phase velocity. The unique free parameter (besides the initial values of the trajectories), is the standard deviation of the Gaussian. In section III, we enlarge the set of initial wave functions by considering initial Gaussians, also centered at the origin, with phase velocity, hence adding a new parameter to the system. It is shown that unitary evolution of such initial wave functions continue to yield symmetric quantum bounces. As unitary evolution is not a mandatory requirement for mini-superspace wave functions in the de Broglie-Bohm theory, we gave up with unitarity, obtaining, in this way, asymmetric quantum bounces. In section IV, we enlarge once more the class of initial wave functions by taking superpositions with two more free parameters than the standard deviation of the Gaussian, obtaining asymmetric quantum bounces with unitarity preserved. In the Conclusion, we comment on our results, and discuss future developments.

\section{De Broglie-Bohm quantization of the mini-superspace Friedmann model}

For a flat, homogeneous and isotropic universe filled with a perfect fluid with equation of state $P=\omega \rho$, where $P$ is the pressure, $\rho$ is the energy density and $\omega$ is the equation of state parameter, the ADM \cite{Arnowitt:1962hi} and the Schutz \cite{Schutz:1970my} formalisms lead to the following Hamiltonian
\begin{equation}
\label{hamiltonian}
H=\frac{L_p^2}{V} N H_{0},
\end{equation}   
with
\begin{equation}
\label{hamiltonian constraint}
H_{0} \equiv \frac{P_{T}}{a^{3 \omega}}-\frac{P_{a}^{2}}{4a},
\end{equation}   
where $L_p$ is the Planck length, $V$ is the volume of the co-moving homogeneous 3-dimensional hyper-surface, which we are supposing to be compact, $a$ is the scale factor of the universe, $T$ is the parameter related to the degree of freedom of the fluid, which plays the role of time, $P_{a}$ and $P_{T}$ are their respective canonically conjugated momenta, and $N$ is the lapse function. We are using natural units, $\hbar=c=1$, hence all canonical variables above are dimensionless, and the Hamiltonian has dimensions of energy $=1/$length, as it should be. The constant $L_p/V$ will be absorbed in the definition of time later on, yielding a dimensionless cosmic time\footnote{This result is obtained from the Einstein-Hilbert action written in terms of the ADM and the Schutz formalisms. One can perform the Legendre transformation in order to find the Hamiltonian density, integrate in the spatial coordinates, and implement a canonical transformation in the fluid variables, leading to Eq.~(\ref{hamiltonian}). The factor $1/4$ comes from the gravitational part of the action, more specifically from the relation between $\dot{a}$ and the conjugated momentum $P_{a}$.}. The Friedmann equations can be readily obtained from the Hamiltonian
\begin{equation}
H=NH_{0},
\end{equation}
where N is the lapse function of the ADM formalism. Applying the Dirac quantization procedure for constrained systems, where the wave function is annihilated by the the constraint operator, $\hat{H_0}\Psi=0$, and taking into account a particular choice of the factor ordering \cite{Halliwell}, which leads to a Schr\"{o}dinger equation with a covariant Laplacian under redefinitions of $a$, we arrive at the following Wheeler-DeWitt equation: 
\begin{equation}\label{wdw equation in a}
i \frac{\partial}{\partial T} \Psi(a,T)=\frac{a^{(3 \omega -1)/2}}{4} \frac{\partial}{\partial a} \left [ a^{(3 \omega -1)/2} \frac{\partial}{\partial a} \right ] \Psi(a,T).
\end{equation}
Performing the variable transformation given by
\begin{equation}\label{chi and a}
\chi =\frac{2}{3(1-\omega)} a^{3(1-\omega)/2},
\end{equation}
we obtain 
\begin{equation}\label{wdw equation}
i \frac{\partial \Psi(\chi,T)}{\partial T}=\frac{1}{4} \frac{\partial^{2}\Psi(\chi,T)}{\partial \chi^{2}},
\end{equation}
which can be identified as a Sch\"{o}dinger equation for a free particle of mass $m=2$ in one dimension with the opposite sign of the time derivative term. The solutions of Eq.~\eqref{wdw equation} are the wave functions of the universe. With the choice $N=a^{3\omega}$ for the lapse function, the parameter $T$ relates to the dimensionless cosmic time $t=(L_p^2/V)t_c$ through $dt=a^{3\omega}dT$, where $t_c$ is the usual cosmic time, with dimension of length.

Once the scale factor $a$ and, consequently, the variable $\chi$ must assume positive values, we are dealing with a Schr\"{o}dinger equation for a particle with negative kinetic energy in the half axis \cite{Gitman}. In order to obtain unitary solutions and, as a consequence, a consistent probabilistic interpretation, it is necessary to perform
a self-adjoint extension, that is, to consider the perfectly reflecting boundaries, which are given by the following condition:
\begin{equation}\label{boundary condition}
\left ( \Psi^{*} \frac{\partial \Psi}{\partial \chi}-\Psi \frac{\partial \Psi^{*}} {\partial \chi} \right )\Biggr{|}_{\chi=0} = 0.
\end{equation}
Note, however, that the de Broglie-Bohm quantum theory is a dynamical fundamental theory, where probabilities arise in a secondary step, as in Classical Mechanics. And indeed, a probabilistic interpretation of the wave function of the Universe may not make sense, since there is only one universe in this approach. A probabilistic interpretation is required only for subsystems in the Universe, where we can perform measurements. In this situation, one can use the so called conditional wave functions for subsystems, in which the Wheeler-DeWitt equation reduces to an unitary Schr\"{o}dinger form, and a probabilistic interpretation where the Born rule is valid can be recovered, which is called quantum equilibrium, see Ref.~\cite{Falciano:2008nk} for details. Of course this opens the possibility that during this process violations of standard quantum mechanics might occur. Unfortunately,
almost all systems in Nature have evolved to the quantum equilibrium phase, where the probability
distribution is described by $\rho$, see Refs.~\cite{Val1,Val2} for detailed investigations about this process, and possible exceptions. Concluding, in what follows, we will not require unitary evolution as necessary feature of the mini-superspace wave function. 

Writing the wave function as $\Psi=Re^{iS}$, and substituting into 
Eq.~(\ref{wdw equation in a}), we obtain two real equations,

\begin{eqnarray}\label{continuity equation}
 &&\frac{\partial \rho}{\partial T}-\frac{\partial}{\partial a} \left[ \frac{a^{(3\omega-1)}}{2} \frac{\partial S}{\partial a} \rho \right] =0\\ \label{hamilton jacobi equation}
\nonumber &&\frac{\partial S}{\partial T}-\frac{a^{(3\omega-1)}}{4} \left( \frac{\partial S}{\partial a}\right)^{2}\\ &&+\frac{a^{(3\omega-1)/2}}{4R} \frac{\partial}{\partial a} \left[ a^{(3\omega-1)/2} \frac{\partial R}{\partial a} \right]=0,
\end{eqnarray}
where $\rho(a,T)=a^{(1-3\omega)/2}|\Psi|^{2}$.

The key feature of the de Broglie-Bohm quantum theory is to assume that positions in configuration space (in our case $a$) have objective reality, independently of any observation, and satisfy the so called guidance equation
\begin{equation}\label{trajectory equation}
\dot{a}=-\frac{a^{(3\omega-1)}}{2} \frac{\partial S}{\partial a},
\end{equation} 
or 
\begin{equation}\label{trajectory equation in chi}
    \frac{d\chi}{dT}=-\frac{1}{2}\frac{\partial S}{\partial \chi} .
\end{equation}

With Eq.~\eqref{trajectory equation}, one can interpret Eq.~\eqref{continuity equation} as a continuity equation for the distribution $\rho$, and Eq.~\eqref{hamilton jacobi equation} as a generalized Hamilton-Jacobi equation
supplemented by the so called quantum potential,
\begin{equation}
Q \equiv -\frac{a^{(3\omega-1)/2}}{4R}\frac{\partial}{\partial a} \left[ a^{(3\omega-1)/2} \frac{\partial R}{\partial a} \right].
\end{equation} 
If one wants to recover the physical dimensions of Eqs.~\eqref{continuity equation} and \eqref{hamilton jacobi equation}, one can easily verify that Planck constant $\hbar$ re-appears only multiplying the quantum potential, $Q\rightarrow \hbar^2 Q$. Hence $Q$ brings the quantum effects to the dynamics. Once the total energy given by Eq.~\eqref{hamilton jacobi equation} includes also the quantum potential $Q$, the trajectory given by Eq.~\eqref{trajectory equation} will not be the same as the classical one, unless $Q$ is negligible with respect to the other terms. This effect is responsible for the emergence of the quantum bounce, avoiding the standard classical initial singularity.

Let us consider an initial wave function of the universe given by
\begin{equation}\label{symmetric wave function}
\Psi_{0}(\chi)=\left( \frac{8}{\pi\sigma^{2}}\right)^{\frac{1}{4}} \exp \left( -\frac{\chi^{2}}{\sigma^{2}} \right),
\end{equation}
which satisfies the boundary condition \eqref{boundary condition}.
In order to obtain an unitary evolution, we must apply the correspondent propagator to the Wheeler-DeWitt equation \eqref{wdw equation} considering the boundary condition \eqref{boundary condition}. It means that we must sum two propagators of a Schr\"{o}dinger equation with negative kinetic energy, one to $\chi_{0}$ and another to $-\chi_{0}$. We then obtain
\begin{eqnarray}\label{propagator}
\nonumber G(\chi,\chi_{0},T)&=&\sqrt{-\frac{i}{\pi T}} \exp \left[-\frac{i (\chi-\chi_{0})^{2}}{T}\right]\\
 &+&\sqrt{-\frac{i}{\pi T}} \exp\left[-\frac{i (\chi+\chi_{0})^{2}}{T}\right].
\end{eqnarray}
The propagator (\ref{propagator}) is not the most general one that satisfies the boundary condition (\ref{boundary condition}). One could, for instance, change the relative sign to minus in order to obtain $G(\chi=0)=0$. However, this propagator leads to a trivial solution for the propagated wave function of the universe. Thus, in practice, the propagator that results in a non-trivial solution satisfies a more restrictive boundary condition, which is given by the von Neumann condition $\partial_{\chi}G|_{\chi=0}=0$. Superpositions of the propagators with relative signs plus and minus with a phase difference of $\pm \pi/2$ are also allowed. However, the only difference in the propagated wave function is a factor that does not modify the Bohmian trajectories.

Applying \eqref{propagator} to the initial wave function (\ref{symmetric wave function}), we arrive at the wave function for all times 
\begin{eqnarray}
&& \nonumber \Psi(\chi,T)=\left[ \frac{8 \sigma^{2}}{\pi(\sigma^{4}+T^{2})} \right]^{\frac{1}{4}} \exp \left[-\frac{\sigma^{2}\chi^{2}}{\sigma^{4}+T^{2}}\right]\\
 &&\times \exp\left[ - i \left(\frac{T \chi^{2}}{\sigma^{4}+T^{2}}+\frac{1}{2} \arctan \left(\frac{\sigma^{2}}{T}\right)-\frac{\pi}{4}\right)\right],
\end{eqnarray}
which also satisfies Eq.~\eqref{boundary condition}.
Using the phase $S$ of the above wave function, we are able to obtain the trajectory of the parameter $\chi$ through Eq.~\eqref{trajectory equation in chi}. It reads
\begin{equation}\label{symmetric bounce in chi}
\chi(T)=\chi_{b}\left[ 1+\left( \frac{T}{\sigma^{2}} \right)^{2} \right]^{\frac{1}{2}},
\end{equation}
where $\chi_{b}$ is the value of $\chi$ at the bounce, which occurs at $T=0$. One can re-obtain the classical solution by taking a Gaussian infinitely peaked. In order to do that, one should consider the differential equation with initial condition $\chi_{0}=\chi(T_{0})$, which leads to the solution
\begin{equation}
    \chi(T)=\chi_{0}\frac{\sqrt{T^{2}+\sigma^{4}}}{\sqrt{T_{0}^{2}+\sigma^{4}}}.
\end{equation}
Then, by making $\sigma^{2} \rightarrow  0$, the classical cosmology given by $\chi(T)=\chi_{0}T/T_{0}$ is obtained.

\begin{figure}[h]
    \centering
    \includegraphics[scale=0.9]{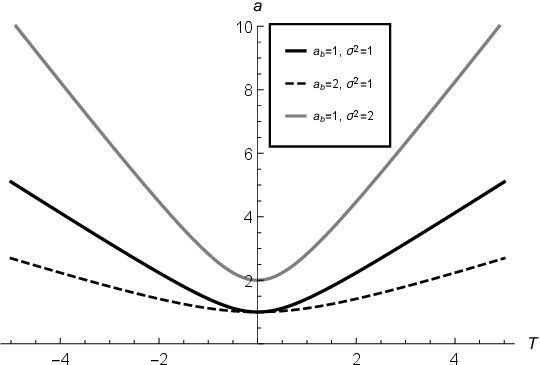}
    \caption{$a$ vs $T$ for  $\omega=\frac{1}{3}$.}
    \label{fig:bounce simetrico}
    \end{figure}

In terms of the scale factor $a$ one gets,
\begin{equation}\label{symmetric bounce in a}
a(T)=a_{b}\left[ 1+\left( \frac{T}{\sigma^{2}} \right)^{2} \right]^{\frac{1}{3(1-\omega)}},
\end{equation}
where $a_{b}$ and $\chi_{b}$ are related also through Eq.~\eqref{chi and a}.
Eq.~\eqref{symmetric bounce in a} describes a symmetric bounce, which is plotted in figure \ref{fig:bounce simetrico}. It tends to the classical solution for large values of $T$.

A good model for the perfect hydrodynamical fluid in the early universe, where all particles are highly relativistic, is a radiation fluid with $w=1/3$, which will be considered from now on. Note that, in this case, $T=\eta$, the conformal time (remember the relation of $T$ with cosmic time $t$, $dt=a^{3w}dT$).   

    It is convenient to express the bounce solution in terms of cosmological quantities, which is achieved by relating the parameters of the wave function to observables. With this purpose, we will follow the same procedure developed in \cite{Celani2017}. We first obtain the Hubble function, given by $H=\frac{\dot{a}}{a}$, where dot denotes the derivative with respect to the physical cosmic time\footnote{When relating the parameters with cosmological observables, one must go back to the physical cosmic time, $t_c=(V/L_p^2)t$. The constant $V/L_p^2$ can be absorbed in the dimensionless variance $\sigma$, see Eq.~\eqref{symmetric bounce in chi}, yielding a variance with dimensions of ${\rm length}^{1/2}$. This turns the subsequent equations with the correct physical dimensions.}. We then take an expansion of the Hubble function squared for large times $T$, which reads
    \begin{equation}\label{hubble2 simetrico}
H^{2}=\frac{a_{b}^{2}}{a^{4}\sigma^{4}}=H_0^2\Omega_{r0}\frac{a_0^4}{a^4},
    \end{equation}
where in the last equality we used the classical Friedmann equation, yielding
\begin{equation}\label{omegar0 simetrico}
        \Omega_{r0}=\frac{a_{b}^{2}}{a_{0}^{4}H_{0}^{2}\sigma^{4}},
\end{equation}\\
where $\Omega_{r0}=\rho_{r0}/\rho_{c0}$ is the dimensionless density parameter for radiation today. The subscript $_{0}$ in all quantities indicates their current values. The quantities $\rho_{r0}$ and $\rho_{c0}=3 H_{0}^{2}/8\pi G$ are, respectively, the current energy density of radiation and the current critical density.
    
    Performing the following transformation of variables
    \begin{eqnarray}\label{tc simetrico}
        x_{b}&=&\frac{a_{0}}{a_{b}}\\
        \overline{\sigma}&=&\sigma \sqrt{a_{0}H_{0}},
    \end{eqnarray}
    we obtain
    \begin{equation}\label{sigma2 simetrico}
        \overline{\sigma}^{2}=\frac{1}{x_{b} \sqrt{\Omega_{r0}}}.
    \end{equation}
    In its turn, the curvature scale at the bounce is given by
    \begin{equation}\label{Lb simetrico}
        L_{b}=\frac{1}{\sqrt{R}}\bigg. \bigg|_{T=0}=\frac{\overline{\sigma}^{2}}{\sqrt{6}x_{b}H_{0}}=\frac{1}{x_{b}^{2}H_{0}\sqrt{6 \Omega_{r0}}},
    \end{equation}
    where $R$ is the Ricci scalar.
    
     To ensure that the Wheeler-DeWitt equation is a valid approximation for a more fundamental theory of quantum gravity \cite{Kiefer}, we must require that the bounce scale is larger than the Planck scale, that is $L_{b}>L_{p}$. Taking $H_{0} \approx  70 \ km \times s^{-1}\times Mpc^{-1}$, $\Omega_{r0} \approx 10^{-4}$ and given that $L_{p}/R_{H0}\approx 1.25 \times 10^{-61}$, where $R_{H0}=1/H_0$ is the Hubble radius today, we obtain the upper bound for $x_{b}$
    
    \begin{equation}
        x_{b}<1.8 \times 10^{31}.
    \end{equation}
    
    The lower limit can be obtained by requiring that the bounce occurs at energy scales much larger than the nucleosynthesis energy scale, i.e. $T_{BBN}=10$ MeV. Using the CMB temperature equal to $T_{\gamma0}=2.7 \ K$ in Mev, and the linear relation between the temperature and the scale factor
    \begin{equation}
       \frac{T_{\gamma0}}{T_{BBN}}=\frac{a_{BBN}}{a_{0}}=x_{BBN}^{-1}, 
    \end{equation}
    we obtain
    \begin{equation}
        x_{b} \gg 10^{11}.
    \end{equation}

\section{\label{sec:citeref}Generalized symmetric bounces and non-unitary asymmetric bounces}
\subsection{\label{sec:Generalized symmetric quantum bounces}Generalized symmetric quantum bounces}
Although the simplicity of the previous symmetric bounce, it represents a fine-tuning in the theory, since the contraction phase is restricted to be the same as the expansion reversed in time. For this reason, we aim to obtain cosmological models with asymmetric trajectories for the scale factor $a$. 

Our initial proposal to obtain asymmetric solutions was to include a factor of the form $\exp(ip\chi)$ in the initial wave function, which represents a velocity for the Gaussian proposed in Eq.~\eqref{symmetric wave function}. Thus we have 
\begin{equation}\label{non unitary initial wf}
   \Psi_{0}(\chi)=\left( \frac{8}{\pi\sigma^{2}}\right)^{\frac{1}{4}} 
    \exp \left( -\frac{\chi^2}{\sigma^2}+ip \chi    \right).
\end{equation}
Note that this initial wave function does not satisfy the boundary condition (\ref{boundary condition}), which means that unitarity is not satisfied at $T=0$. However, implementing a convolution between this initial wave function and a propagator that satisfies condition (\ref{boundary condition}), we are, in practice, dealing with the projection of $\Psi_{0}$ onto the subspace of square-integrable functions on the $\chi$ half-line satisfying the von Neumann boundary condition. As a result, the propagated wave function that results from this convolution is going to satisfy (\ref{boundary condition}).

Propagating this initial wave function \eqref{non unitary initial wf} with the propagator \eqref{propagator} from $0$ to $+\infty$, that is, performing a unitary evolution, we obtain the following wave function for all times:
\begin{eqnarray}\label{wf nu to u}
  \nonumber  \Psi(\chi,T)&=&(2 \pi \sigma^{2})^{-\frac{1}{4}}\biggl(-1+\frac{iT}{\sigma^{2}} \biggl)^{-\frac{1}{2}}\\
  &\times& \biggl[ \phi(\chi,T)+\phi(-\chi,T) \biggr] ,
\end{eqnarray}
where 
\begin{eqnarray}
    \nonumber  \phi(\chi,T)& \equiv &\exp \biggl[ -\frac{\sigma^{2}\chi^{2}}{T^{2}+\sigma^{4}}-\frac{T(p^{2}T\sigma^{2}-4p\sigma^{2}\chi)}{4(T^{2}+\sigma^{4})}\\
     \nonumber &+&i \biggl( -\frac{T\chi^{2}}{T^{2}+\sigma^{4}}+\frac{\sigma^{2}(p^{2}T\sigma^{2}-4p\sigma^{2}\chi)}{4(T^{2}+\sigma^{4})} \biggr) \biggr]\\
     &\times& \biggl( 1- \operatorname { Erf } \left[ \epsilon(\chi,T) \right] \biggr)
\end{eqnarray}
and 
\begin{equation}
    \epsilon(\chi,T) \equiv  \biggl( \frac{pT}{2}  +  \chi \biggr)   \biggl[ iT \biggl( -1+\frac{iT}{\sigma^{2}} \biggr) \biggr]^{-\frac{1}{2}}.
\end{equation}
The wave function \eqref{wf nu to u} satisfies the boundary condition \eqref{boundary condition}. Thus, as mentioned before, the non-unitarity at the point $T=0$ for the initial wave function \eqref{non unitary initial wf} does not spoil the unitarity after the convolution with the propagator \eqref{propagator}.

We can see from Eq.~\eqref{wf nu to u} that the wave function was propagated equally to $\chi$ and to $-\chi$.  Thus terms and arguments that are linear in $\chi$ are symmetrized with respect to $\chi=0$ by the unitary evolution with the propagator \eqref{propagator}. 

In order to exemplify a Bohmian trajectory for the scale factor $a$ related to an unitary wave function with factors of the form $\exp(ip\chi)$, we are going to consider only the terms
\begin{eqnarray}\label{wf nu to u part}
   \label{psi uni symmetrized}
    \nonumber \overline{\Psi}(\chi,T)&=&(2 \pi \sigma^{2})^{-\frac{1}{4}}\biggl(-1+\frac{i T}{\sigma^{2}} \biggl)^{-\frac{1}{2}} \\
    &\times& \biggl[ \overline{\phi}(\chi,T)+\overline{\phi}(-\chi,T) \biggr],
\end{eqnarray}
where
\begin{eqnarray}\label{phibarra}
    \nonumber  \overline{\phi} (\chi,T)& \equiv &\exp \biggl[ -\frac{\sigma^{2}\chi^{2}}{T^{2}+\sigma^{4}}-\frac{T(p^{2}T\sigma^{2}-4p\sigma^{2}\chi)}{4(T^{2}+\sigma^{4})}\\
     &+&i \biggl( -\frac{T\chi^{2}}{T^{2}+\sigma^{4}}+\frac{\sigma^{2}(p^{2}T\sigma^{2}-4p\sigma^{2}\chi)}{4(T^{2}+\sigma^{4})} \biggr) \biggr],
\end{eqnarray}
which also constitutes a unitary solution of the Wheeler-DeWitt equation \eqref{wdw equation}. The choice to disregard the Gauss's error functions is for the sake of simplicity.

\begin{figure}[h]
    \centering
    \includegraphics[scale=0.9]{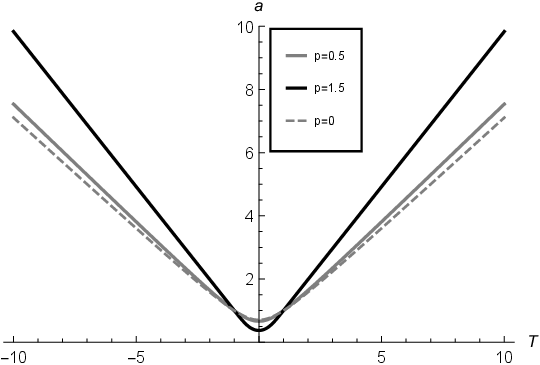}
    \caption{$a$ vs $T$ for $\sigma=1.0$, $a_{i}=1.0$, $T_{i}=1.0$, $\omega=\frac{1}{3}$.}
    \label{fig:nu to u}
    \end{figure}

Inserting the global phase $\overline{S}$ of the wave function \eqref{wf nu to u part} into Eq.~\eqref{trajectory equation in chi}, it is possible to obtain a differential equation for the parameter $\chi$. It reads
\begin{widetext}
\begin{equation}\label{diff eq nu to u}
 \frac{d\chi}{dT}=\frac{2T\chi\cos \biggl ( \frac{2p\sigma^{4}\chi}{T^{2}+\sigma^{4}} \biggr ) +2T\chi\cosh \biggl( \frac{2pT\sigma^{2}\chi}{T^{2}+\sigma^{4}} \biggr)+pT\sigma^{2}\sin\biggl ( \frac{2p\sigma^{4}\chi}{T^{2}+\sigma^{4}} \biggr )+p\sigma^{4} \sinh \biggl( \frac{2pT\sigma^{2}\chi}{T^{2}+\sigma^{4}} \biggr)}{2(T^{2}+\sigma^{4})\biggl[ \cos\biggl ( \frac{2p\sigma^{4}\chi}{T^{2}+\sigma^{4}} \biggr ) +\cosh \biggl( \frac{2pT\sigma^{2}\chi}{T^{2}+\sigma^{4}} \biggr)\biggr]}.
\end{equation}

\end{widetext}
Using Eq.~\eqref{chi and a} in Eq.~\eqref{diff eq nu to u} and solving it numerically with initial condition $a_{i}=a(T_{i})$, we obtain the trajectory of the scale factor $a(T)$, which is plotted in figure \ref{fig:nu to u}.

The result is a symmetric bounce, regardless of the value of the parameter $p$ related to the asymmetry. It happens when the unitary evolution for factors of the form $\exp(ip\chi)$ is maintained. As explained before, once these factors are linear in $\chi$ inside the exponential, they are going to be propagated equally to $\chi$ and to $-\chi$, resulting in a symmetrization of the propagated wave function and, as a consequence, of the trajectory of the scale factor $a$. 

Note that different symmetric bounces can be obtained in other approaches to quantum cosmology. For instance, in Refs~\cite{Gryb1,Gryb2}, a relational quantization method was implemented, where unitarity is a necessary requirement in order to obtain a consistent probabilistic interpretation, and bouncing models were also found. On the other hand, our work relies on a deterministic interpretation of quantum mechanics, where probabilities are not fundamental, allowing to explore the consequences of wave functions of the Universe which are not restricted to evolve satisfying unitarity requirements. 

\subsection{\label{sec:Non-unitary asymmetric quantum bounces}Non-unitary asymmetric quantum bounces}

An alternative to this hindrance is to give up unitarity, which is allowed according to the discussion previously made. In practise, it means to disconsider the boundary condition \eqref{boundary condition}.
The correspondent propagator is then only the first term of the propagator \eqref{propagator}, given by 
\begin{equation}\label{nu propagator}
    G^{NU}(\chi,\chi_{0},T)=\sqrt{-\frac{i}{\pi T}} \exp \left[-\frac{i (\chi-\chi_{0})^{2}}{T}\right],
\end{equation}
where $NU$ stands for non-unitary. 
Applying the propagator \eqref{nu propagator} to the initial wave function \eqref{non unitary initial wf} without the normalization factor from $-\infty$ to $+\infty$, we obtain the following wave function for all times:

\begin{small}
\begin{equation}\label{non unitary wf}
   \Psi(\chi,T)= \biggl( -1+\frac{i T}{\sigma^{2}} \biggr)^{-\frac{1}{2}} \exp \left( \frac{\frac{ip^{2}T}{4}+ip\chi-\frac{ \chi^{2}}{\sigma^{2}}}{1-\frac{iT}{\sigma^{2}}}\right)   .
\end{equation}
\end{small}
We take the integration from $-\infty$ to $\infty$ in Eq.~(\ref{nu propagator}) in order to avoid terms containing Gauss error functions that arise if the integration is performed from $0$ to 
$\infty$. In the end we must check that the restriction $\chi>0$ is still staisfied.

\begin{figure}[h]
    \centering
    \includegraphics[scale=0.9]{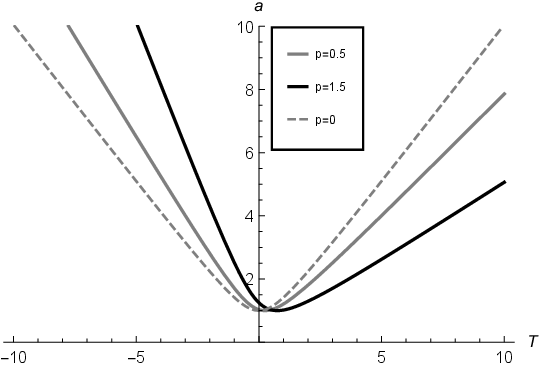}
    \caption{$a$ vs $T$ for $\sigma=1.0$, $a_{b}=1.0$, $\omega=\frac{1}{3}$.}
    \label{fig:bounce assimetrico nu 1}
    \end{figure}

Writing Eq.~\eqref{non unitary wf} as $\Psi(\chi,T)=R(\chi,T)e^{iS(\chi,T)}$, we obtain
\begin{equation}
    \Psi(\chi,T)=\biggr( -1+\frac{iT}{\sigma^{2}}\biggl)^{-\frac{1}{2}} \overline{\phi}(-\chi,T),
\end{equation}
where $\overline{\phi}(\chi,T)$ is given by Eq.~\eqref{phibarra} (the first factor in the above equation does not depend on $\chi$, hence it does not affect the calculation of the Bohmian trajectories).
Then, by inserting $S$ into Eq.~\eqref{trajectory equation in chi}, it is possible to obtain the trajectory in terms of $\chi$. It reads
\begin{equation}\label{chi13}
    \chi(T) = \chi_{b} \biggl[ 1+ \left( \frac{T}{\sigma^{2}} \right) ^{2}+ \left( \frac{p}{2 \chi_{b}} \right) ^{2}(T^{2}+\sigma^{4}) \biggr]^{\frac{1}{2}} -\frac{pT}{2},
\end{equation}
where $\chi_{b}=\chi(T_{b})$ is the value of the variable $\chi$ at the moment of the bounce $T_{b}=\frac{p\sigma^{4}}{2 \chi_{b}}$, which is not equal to zero as in the symmetric case. In terms of the scale factor, the trajectory reads
\begin{eqnarray}\label{nu trajectory}
    \nonumber  a(T) &=& \Biggl \{ \Biggr. -\frac{3p(1-\omega)}{4}T+ a_{b}^{\frac{3(1-\omega)}{2} }
    \Biggl[ 1+ \left( \frac{T}{\sigma^{2}} \right) ^{2}\\
     &+&\left( \frac{3p(1-\omega)}{4} \right) ^{2}  \frac{(T^{2}+\sigma^{4})}{a_{b}^{3(1-\omega)}} \Biggr]^{\frac{1}{2}} \Biggl . \Biggr \} ^{\frac{2}{3(1-\omega)}},
\end{eqnarray}
where $a_{b}$ relates to $\chi_{b}$ through Eq.~\eqref{chi and a}. The trajectory \eqref{nu trajectory} is shown in figure \ref{fig:bounce assimetrico nu 1} for $w=1/3$, where it is evidenced that the value of the parameter $p$ is directly related to the intensity of the asymmetry.

Note that Eq.~\eqref{nu trajectory} does not admit a singularity or negative values for $a(T)$, \textbf{since we always have 
\begin{eqnarray}
   \nonumber &&\frac{3p(1-\omega)}{4}T
   <a_{b}^{\frac{3(1-\omega)}{2}}\\
    &&\times \Biggl[ 1+ \left( \frac{T}{\sigma^{2}} \right) ^{2}
     +\left( \frac{3p(1-\omega)}{4} \right) ^{2}  \frac{(T^{2}+\sigma^{4})}{a_{b}^{3(1-\omega)}}\Biggr]^{\frac{1}{2}}.
\end{eqnarray}}
This ensures that the restrictions $\chi>0$ and $a>0$ are satisfied, although we have disregarded the boundary condition \eqref{boundary condition} and propagated the wave function from $-\infty$ to $\infty$. 
A bounce solution is naturally obtained, without the need to impose restrictions to recover the positivity of the scale factor. 

For $p=0$ we re-obtain the symmetric bounce \eqref{symmetric bounce in a}, which makes explicit the relation between the asymmetry and the factor $\exp(ip\chi)$. 

As in the symmetric case, the classical solution
arises for large values of $T$.

\begin{figure}[h]
    \centering
    \includegraphics[scale=0.9]{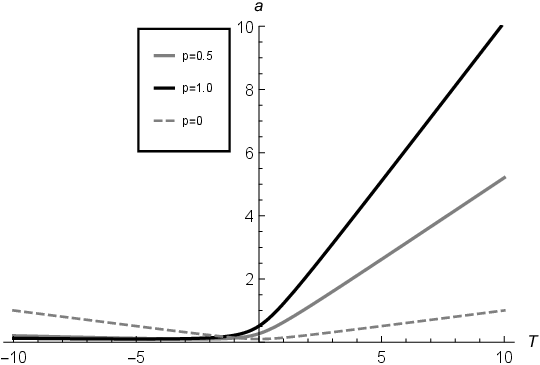}
    \caption{$a$ vs $T$ for $\sigma=1.0$, $a_{b}=0.1$, $\omega=\frac{1}{3}$.}
    \label{fig:bounce assimetrico nu 2}
    \end{figure}

In order to obtain a slope in the contracting phase lower than the slope in the expanding phase, one has to take $p < 0$, or, equivalently, to change the factor from $\exp(ip\chi)$ to $\exp(-ip\chi)$ in the initial wave function \eqref{non unitary initial wf} keeping $p>0$. This case is particularly interesting, once the contraction phase may consist of an almost Minkowski universe. Applying the same procedure to obtain the Bohmian trajectory, we obtain $a$, which is plotted in figure \ref{fig:bounce assimetrico nu 2}.

    Just as we did for the symmetric case, let us express the wave function parameters in terms of cosmological quantities for the case $w=1/3$. Defining the parameters
    \begin{eqnarray}\label{tc assimetrico}
        x_{b}&=&\frac{a_{0}}{a_{b}}\\
        \overline{\sigma}&=&\sigma \sqrt{a_{0}H_{0}}\\
        \overline{p}&=&\frac{p}{a_{0}^{2}H_{0}}\\
        \overline{\eta}&=&\frac{\eta}{\sigma^2}\\
        y^2 &=& \frac{x_b\overline{p}{\overline{\sigma}}^2}{2},
    \end{eqnarray}
    one can write
    \begin{equation}\label{aS}
        a= a_b\biggl(\pm y^2\overline{\eta}+
        \sqrt{1+y^{4}}\sqrt{1+{\overline{\eta}}^2}\biggr),
    \end{equation}
    where the $\pm$ signs correspond to wave function phases $\exp(\mp ip\chi)$, with $p\geq 0$. 
    In the limit $|\overline{\eta}|>>1$, we get for the Hubble function,
    \begin{equation}\label{Hpe}
            H^2=\frac{\biggl(\pm y^2+
        \sqrt{1+y^{4}}\biggr)^2 a_b^2 H_0^2 a_0^2}{{\overline{\sigma}}^4a^4}=H_0^2\Omega_{r0}\frac{a_0^4}{a^4},
    \end{equation}
    in the expanding phase, and
    \begin{equation}\label{Hpc}
            H^2=\frac{\biggl(\mp y^2+
        \sqrt{1+y^{4}}\biggr)^2 a_b^2 H_0^2 a_0^2}{{\overline{\sigma}}^4a^4}=H_0^2\Omega_{rc}\frac{a_0^4}{a^4},
    \end{equation}
    in the contracting phase, where $\Omega_{rc}$ is the radiation energy density when the Universe has $H=H_0$ in the contracting phase divided by the critical density $\rho_c$. These equations imply that
    \begin{equation}\label{omegar0 assimetrico}
    \Omega_{r0} =  
    \frac{\left(\pm y^2+
        \sqrt{1+y^{4}}\right)^2}{{\overline{\sigma}}^4x_b^4} , 
    \end{equation}
    \begin{equation}\label{sigma2 assimetrico}
        \overline{\sigma}^{2}=\left[ x_{b}^{2}\Omega_{r0}\biggl(1\mp \frac{\overline{p}}{\sqrt{\Omega_{r0}}}\biggr)\right]^{-1/2},
    \end{equation}
    and
    \begin{equation}
    \label{Omegarc}
\Omega_{rc}=\Omega_{r0}\biggl(1\mp \frac{\overline{p}}{\sqrt{\Omega_{r0}}}\biggr)^{2}.
    \end{equation}
    
Note that the $+$ sign in Eq.~\eqref{aS} implies, from Eq.~\eqref{sigma2 assimetrico}, that $0\leq \overline{p} < \sqrt{\Omega_{r0}}$. From Eq.~\eqref{Omegarc}, one can see that $\Omega_{rc} \leq \Omega_{r0}$, and in the limit $ \overline{p} \to \sqrt{\Omega_{r0}}$ one has $\Omega_{rc} \to 0$. Hence, the contracting universe can be made arbitrarily flat, and the radiation fluid is created around the quantum phase, during the bounce.
    
In the $-$ sign case in Eq.~\eqref{aS}, there is no constraint in $\overline{p}$, $0\leq \overline{p} < \infty$, and $\Omega_{rc} \geq \Omega_{r0}$.
    
In this asymmetric case, the maximum curvature does not occur at the bounce, ${\overline{\eta}}_{\rm bounce} \mp y^2$, but at the conformal time ${\overline{\eta}}_{\rm max} \mp \sqrt{\frac{\sqrt{1+y^4}-1}{2}}$. Hence, the minimum curvature scale reads 
\begin{eqnarray}\label{Lm assimetrico}
L_{\rm min}&=&\frac{1}{\sqrt{R}}\bigg. \bigg|_{{\overline{\eta}}_{\rm max}}\nonumber\\&=&\frac{R_{H0}\left(1+\sqrt{1\mp\frac{\overline{p}}{\sqrt{\Omega_{r0}}}}\right)^3}{8\sqrt{3\Omega_{r0}}x_{b}^{2}\left(1\mp\frac{\overline{p}}{\sqrt{\Omega_{r0}}}\right)^2\sqrt{\left(2\mp\frac{\overline{p}}{\sqrt{\Omega_{r0}}}\right)}}.
\end{eqnarray}
    Note that Eqs.~(\ref{sigma2 assimetrico}, \ref{Lm assimetrico}) reduce to their correspondents in the symmetric case Eqs.~(\ref{sigma2 simetrico}, \ref{Lb simetrico}) for $\overline{p}=0$.

As in the symmetric case, we require that the bounce scale is larger than the Planck scale, that is $L_{\rm min}>L_{p}$, and smaller then the curvature scale at nucleosynthesis. Hence, we demand

\begin{equation}
\label{cond asym}
10^{-58} << \frac{L_{\rm min}}{R_{H0}} < 10^{-20}.
\end{equation}

Note that, in the asymmetric case, there is no direct relation between $x_b$ and $L_{\rm min}$ due to the presence of $\overline{p}$ in 
Eq.~(\ref{Lm assimetrico}). Hence, neither $x_b$ nor $\overline{p}$ have independent physical significance, just when combined to give $L_{\rm min}$. That is why, in this case, the condition must be put in terms of \eqref{cond asym}.

\section{\label{sec:Unitary asymmetric quantum bounces}Unitary asymmetric quantum bounces}
Another alternative to obtain asymmetric solutions is to perform superpositions of Gaussian wave functions multiplied by factors of the form $\exp[i(p\chi)^{2}]$. Once the term inside the exponential is not linear in $\chi$, it is possible to generate asymmetry maintaining unitarity. Note that the asymmetry is achieved only when we perform superpositions. A single Gaussian in this format would lead to a symmetric bounce. 

Considering the following superposition for the initial wave function
\begin{eqnarray}\label{wf0 unitary asymm}
\nonumber \Psi_{0}(\chi)&=& C \biggl[ \exp \left( -\frac{\chi^{2}}{\sigma^{2}}+i p_{1}^{2} \chi^{2} \right)\\
  &+& \exp \left( -\frac{\chi^{2}}{\sigma^{2}}-i p_{2}^{2} \chi^{2} \right) \biggr] ,
 \end{eqnarray}
where 
\begin{eqnarray}
   \nonumber C&=&\frac{\sqrt{2}}{ \pi^{\frac{1}{4}}} \biggl \{   \left[ -i(p_{1}^{2}+p_{2}^{2})+\frac{2}{\sigma^{2}} \right]^{-\frac{1}{2}} \\
    &+&\left[ i(p_{1}^{2}+p_{2}^{2})+\frac{2}{\sigma^{2}} \right]^{-\frac{1}{2}} +\sqrt{2} \sigma \biggr\}^{-1/2},
\end{eqnarray}
and applying the unitary propagator \eqref{propagator}, we obtain a wave function for all times given by 
\begin{widetext}
\begin{footnotesize}
\begin{eqnarray}\label{wf unitary asymm}
 \Psi(\chi,T)=\frac{
C  \exp \left( -i\frac{\chi^{2}}{T} \right) 
  \biggl \{ \exp \left[ \frac{i  \chi^{2}}{T-iT^{2}(\frac{1}{\sigma^{2}}+ip_{2}^{2})} \right] \biggr( -ip_{1}^{2}+\frac{i}{T}+\frac{1}{\sigma^{2}} \biggl)^{\frac{1}{2}}
  + \exp  \left[  \frac{i  \chi^{2}}{T-iT^{2}(\frac{1}{\sigma^{2}}-ip_{1}^{2})} \right] \biggl( ip_{2}^{2}+\frac{i}{T}+\frac{1}{\sigma^{2}} \biggr)^{\frac{1}{2}} \biggr \}}{\left[ iT \left( -ip_{1}^{2}+\frac{i}{T}+\frac{1}{\sigma^{2}} \right) \left( ip_{2}^{2}+\frac{i}{T}+\frac{1}{\sigma^{2}} \right) \right]^{\frac{1}{2}}}.
\end{eqnarray}
\end{footnotesize}
\end{widetext}
Note that both Eq.~\eqref{wf0 unitary asymm} and Eq.~\eqref{wf unitary asymm} satisfy the boundary condition \eqref{boundary condition}. Thus this case is unitary for all times.

Defining
\begin{eqnarray}
    \gamma_{i}=(-1)^{i}p_{i}^{2}+\frac{1}{T}, \quad
    \beta_{i}=\gamma_{i} ^ { 2 } + \frac { 1 } { \sigma ^ {4}}, 
\end{eqnarray}
\begin{eqnarray}
    \nonumber \alpha&=& \frac{\gamma_{1}}{\beta_{1}} \frac{\chi^ { 2 }}{T ^ { 2 }}
    - \frac {\gamma_{2}}{\beta_{2}}\frac{\chi^{2}}{T ^{ 2 }}
    - \frac{1}{2} \arctan \left(  \gamma_{1} \sigma^{2}   \right)\\ &+& \frac{1}{2} \arctan \left( \gamma_{2} \sigma^{2}  \right) 
\end{eqnarray}
and writing Eq.~\eqref{wf unitary asymm} as $\Psi(\chi,T)=R(\chi,T)e^{iS(\chi,T)}$, we can insert the phase $S$ into Eq.~\eqref{trajectory equation in chi} to obtain the differential equation for the parameter $\chi$, given by
\begin{widetext}
\begin{footnotesize}
\begin{eqnarray}\label{diff eq bounce assimetrico unitario}
   \nonumber \frac{d\chi}{dT}&=&-\biggl \{  \biggr.   \exp \left( - \frac { 2 \chi ^ { 2 } } { \sigma^{2} \beta_{1}  T ^ { 2 } } \right) \left( - T +   \frac{\gamma_{1}}{\beta_{1}}  \right)  2 \beta_{2}^{\frac{1}{2}} \frac{\chi}{T^{2}} +
    \exp \left[ -\left(  \frac { 1 } { \beta_{1} T ^ { 2 } } + \frac {1} { \beta_{2} T ^ { 2 } } \right) \frac{\chi ^ { 2 }}{\sigma^{2}}   \right] 
     \left( \beta_{1} \beta_{2} \right) ^ {   \frac{1}{4} } 
     \nonumber \left[  \frac{2  \cos ( \alpha ) \chi}{ T ^ { 2 }}\left( - 2T + \frac { \gamma_{1}} { \beta_{1} } + \frac { \gamma_{2} } { \beta_{2} } \right) 
     + \frac { 2 \sin ( \alpha ) \chi} {  \sigma^{2}\beta_{2}T ^ { 2 }} \right]\\
      \nonumber &+&\exp \left[- \frac { 2 \chi ^ { 2 } } {  \sigma^{2} \beta_{2}T ^ { 2 } } \right] 
     \left[ -\frac { 2 \beta_{1} ^ { \frac{1}{2} } \left( \beta_{2} -\frac{\gamma_{2}}{T} \right) \chi } { \beta_{2} T} -   \exp \left[  -  \left( \frac { 1 } {\beta_{1} T ^ { 2 } }- \frac{1} {\beta_{2} T^{2}} \right) \frac{\chi^{2}}{\sigma^{2}}     \right] \frac{2 \beta_{2}^{\frac{1}{4}}   \sin(\alpha) \chi} { \sigma^{2} \beta_{1}^{\frac{3}{4}} T^{2}} \right] \biggl.  \biggr \} \\
   &\times& \label{dchidT} \biggl \{  \biggr. 2\exp \left( -\frac {2 \chi ^ { 2 } } {  \sigma^{2}\beta_{2} T^{2}} \right)  \beta_{1} ^{\frac{1}{2}} + 2 \exp \left( - \frac { 2 \chi ^ { 2 } } {  \sigma^{2} \beta_{1}T ^ { 2 }}  \right) \beta_{2}^{\frac{1}{2}} +4 \exp \left[ - \left( \frac{1}{\beta_{1}T^{2} }+\frac{1}{\beta_{2}T^{2}} \right) \frac{\chi^{2}}{\sigma^{2}} \right]
    ( \beta_{1} \beta_{2}) ^ {\frac{1}{4}} \cos ( \alpha ) \biggl.  \biggr \} ^{-1}.
\end{eqnarray}
\end{footnotesize}
\end{widetext}
For $p_{1}=0$ and $p_{2}=0$, i.e. $\gamma_{1}=\gamma_{2}=1/T$ and $\beta_{1}=\beta_{2}=1/T^{2}+1/\sigma^{4}$, we obtain
\begin{equation}
    \frac{d\chi}{dT}=\frac{T\chi}{T^{2}+\sigma^{4}},
\end{equation}
which can be solved analytically and results in the trajectory \eqref{symmetric bounce in chi} obtained before for the symmetric case. 

\begin{figure}[h!]
    \centering
    \includegraphics[scale=0.9]{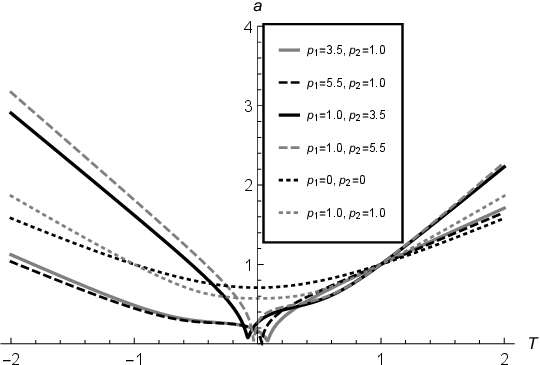}
    \caption{$a$ vs $T$ for $\sigma=1.0$, $a_{i}=1.0$, $T_{i}=1.0$ $\omega=\frac{1}{3}$.}
    \label{fig:bounce assimetrico u 1}
    \end{figure}

Solving Eq.~\eqref{dchidT} numerically with initial condition $a_{i}=a(T_{i})$, we obtain the trajectory for the parameter $\chi$ and then, using Eq.~\eqref{chi and a}, for the scale factor $a$. The result is plotted in figure \ref{fig:bounce assimetrico u 1}. 
Note that symmetric bounces are also obtained if $p_{1}=p_{2}$. 

\begin{figure}[h!]
    \centering
    \includegraphics[scale=0.9]{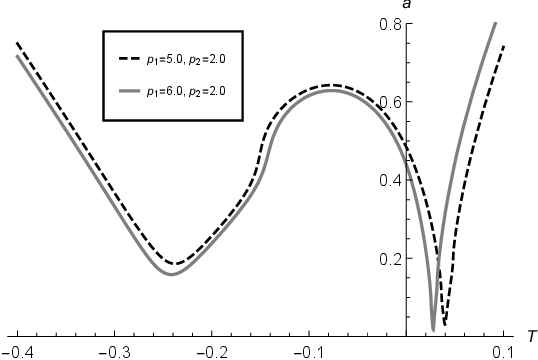}
    \caption{$a$ vs $T$ for $\sigma=1.5$, $a_{i}=5.0$, $T_{i}=1.0$, $\omega=\frac{1}{3}$.}
    \label{fig:multiple bounces}
    \end{figure}

    The numerical solution of Eq.~\eqref{dchidT} also encompasses multiple bounces for certain values of the parameters $\sigma$, $p_{1}$ and $p_{2}$ and of the initial values $a_{i}$ and $T_{i}$. See figure \ref{fig:multiple bounces}.
    
    As we did for the other bounce solutions, we express the wave function parameters in terms of cosmological quantities. Expanding the square of the correspondent Hubble function for large times $T$, we obtain
    \begin{equation}\label{hubble2 assimetrico unitario}
        H^{2}=\frac{a_{i}^{2}}{a^{4}(T_{i}^{2}+\sigma^{4})}.
    \end{equation}
    Identifying the dimensionless density parameter for radiation today $\Omega_{r0}=\rho_{r0}/\rho_{c0}$ as the coefficient of $(a_{0}/a)^{4}$, we obtain
    \begin{equation}\label{omegar0 assimetrico unitario}
        \Omega_{r0}=\frac{a_{i}^{2}}{a_{0}^{4}H_{0}^{2}(T_{i}^{2}+\sigma^{4})}.
    \end{equation}
    
    In order to rewrite Eq.~\eqref{omegar0 assimetrico unitario} in terms of $a_{b}$ and $T_{b}$, we expand Eq.~\eqref{diff eq bounce assimetrico unitario} for $T/\sigma^{2}\ll 1$ to the first order and for $p_{1} \sigma \ll 1$ and $p_{2} \sigma \ll 1$ to the second order. Under these conditions, i.e. near the bounce and with small parameters related to asymmetry, we obtain a solution with a single bounce, where it is possible to relate $T_{b}$, $p_{1}$ and $p_{2}$ by making $da/dT=0$. Disregarding also terms containing $p_{1}^{2}p_{2}^{2}$, we obtain
    \begin{equation}\label{eq tbab p1 and p2}
    T_{b}=\frac{(p_{1}^{2}-p_{2}^{2})\sigma^{4}}{2}.
\end{equation}

Performing the following transformation of variables
    \begin{eqnarray}\label{tc assimetrico unitario p1 e p2}
    x_{b}&=&\frac{a_{0}}{a_{b}}\\
    \overline{\sigma}&=&\sigma \sqrt{a_{0}H_{0}}\\
    \overline{p}_{1}^{2}&=&\frac{p_{1}^{2}}{a_{0}H_{0}}\\
    \overline{p}_{2}^{2}&=&\frac{p_{2}^{2}}{a_{0}H_{0}},
\end{eqnarray}
    we obtain
    \begin{equation}\label{sigma2 assimetrico unitario p1 e p2}
    \overline{\sigma}^{2}=\sqrt{\frac{2}{x_{b}^{2} \Omega_{r0}+\sqrt{x_{b}^{2}\Omega_{r0}[(\overline{p}_{1}^{2}-\overline{p}_{2}^{2})^{2}+x_{b}^{2}\Omega_{r0}]}}}.
\end{equation}
    Note that Eqs.~(\ref{hubble2 assimetrico unitario}, \ref{omegar0 assimetrico unitario}, \ref{sigma2 assimetrico unitario p1 e p2}) reduce to their correspondents in the symmetric case Eqs.~(\ref{hubble2 simetrico}, \ref{omegar0 simetrico}, \ref{sigma2 simetrico}) for $\overline{p}_{1}=\overline{p}_{2}=0$, which implies $T_{i}=T_{b}=0$.
    
     For this particular case, i.e. $T/\sigma^{2}\ll 1$ to first order and for $p_{1} \sigma \ll 1$ and $p_{2} \sigma \ll 1$ to second order, the curvature scale at the bounce $L_{b}$ assumes the same form of the symmetric case given by Eq.~\eqref{Lb simetrico}, but with $\overline{\sigma}^{2}$ given by \eqref{sigma2 assimetrico unitario p1 e p2}. 
     
     We now go back to the general case given by Eq.~\eqref{diff eq bounce assimetrico unitario} and verify for which values of the parameters the bounce scale is larger than the Planck scale and smaller than the nucleosynthesis scale. We find $L_{b}$ numerically for some non-multiple asymmetric bounces, and we obtain the correspondent bounce energy $E_{b}=L_{b}^{-1/2}$ for each case. The results are shown in table \ref{Table1}.
    
\begin{table}[h!]    
\centering
\begin{tabular}{|c|c|c|c| }
\hline
$p_{1}\sigma$ & $p_{2}\sigma$ & $L_{b}$ (s) & $E_{b}$ (MeV)  \\
\hline
 $2.5$ & $1.0$ & $3.59934\times 10^{-3}$ & $16.66820$  \\ 
 \hline
 $3.5$ & $1.0$ & $5.95604 \times 10^{-4}$ & $40.97522$  \\ 
 \hline
 $4.5$ & $1.0$ & $1.61263\times 10^{-4}$ & $78.74681$  \\ 
 \hline
 $5.5$ & $1.0$ & $5.75934 \times 10^{-5}$ & $131.76909$  \\ 
 \hline
 $6.5$ & $1.0$ & $1.19055 \times 10^{-5}$ & $201.63933$  \\ 
 \hline
 $7.5$ & $1.0$ & $4.78629 \times 10^{-5}$ & $289.81846$  \\ 
 \hline
 $8.5$ & $1.0$ & $6.32385 \times 10^{-6}$ & $397.65741$  \\ 
 \hline
 $9.5$ & $1.0$ & $3.60849 \times 10^{-6}$ & $526.42560$  \\ 
 \hline
 $10.5$ & $1.0$ & $2.17979 \times 10^{-6}$ & $677.31783$  \\ 
 \hline
 $1.0$ & $2.5$ & $5.64555 \times 10^{-3}$ & $13.30904$  \\ 
 \hline
 $1.0$ & $3.5$ & $1.00531 \times 10^{-3}$ & $31.53917$  \\ 
 \hline
 $1.0$ & $4.5$ & $2.75388 \times 10^{-4}$ & $60.25975$  \\ 
 \hline
 $1.0$ & $5.5$ & $9.80995 \times 10^{-5}$ & $100.96402$  \\ 
 \hline
\end{tabular}
\caption{$L_{b}$ and $E_{b}$ for $\sigma=1.0, a_{i}=1.0, T_{i}=1.0, \omega=\frac{1}{3}$.} \label{Table1}
\end{table}

Once $L_{p}\approx  5 \times 10^{-44} \ s$, we see that $L_{b}>> L_{p}$ for all bounces considered. As mentioned before, this means that the validity of the Wheeler-DeWitt equation as an approximation to a more fundamental theory of gravity is well established. Beyond that, the bounce must occur at energy scales much larger than the nucleosynthesis scale, i.e. $10$MeV, which is not achieved by all cases considered. Indeed, as one can see from table I, the energy scale of such bounces are not much bigger than the nucleosynthesis energy scale, but they are many orders of magnitude smaller than the Planck energy scale. Hence, the physically relevant consistency check of such bouncing models is the upper limit of $L_b$, not its lower limit, which makes the distinction between $L_b$ and $L_{\rm min}$ irrelevant.

The cases $p_{1}\sigma\geqslant 10.9$, $p_{2}\sigma=1.0$ and $p_{1}\sigma=1.0$, $p_{2}\sigma\geqslant 5.8$ represent multiple bounces. 
Multiple bounces are also encountered in quantum reduced loop cosmology, in a scenario called emergent bounce \cite{LQC}. It describes a series of bounces with successive increasing amplitudes. In our work, the multiple bounces do not necessarily present this behaviour. The solutions we found also allow for more than one bounce, but with similar amplitudes, before being launched to the expanding phase.

\section{Conclusion}

We have obtained generalizations of the quantum bounce solutions obtained in 
Refs~\cite{Pinto-Neto:2013toa,PintoNeto:2004uf} which are asymmetric with respect to the bounce, and even possessing multiple bounces. These solutions may be used to take into account significant back-reaction due to quantum particle production around the bounce, see Refs.~\cite{Celani2017,Scardua2018}. As an example, in future work we will investigate baryogenesis in those asymmetric bounces.

One particular class of interesting solutions is the one exhibited in figure \ref{fig:bounce assimetrico nu 2}. It describes expanding cosmological solutions arising from an almost flat space-time. As discussed in Section III, the energy density at contraction can be made arbitrarily small, depending on the new quantum parameter $p$, related to the phase velocity of the initial wave function of the universe. The emerging picture is of an arbitrarily flat and almost empty space-time, which is launched through a bounce into the standard Friedmann expanding phase, containing the usual hot and dense radiation field. This fact open new windows to an old speculation, that our Universe arose from quantum fluctuations of a fundamental quantum vacuum. The de Broglie-Bohm theory allows a different regard to this hypothesis and the concrete possibility to extend this particular mini-superspace model by incorporating quantum cosmological perturbations to the system and quantitatively study their observational effects. This is also subject for future work. 

\begin{acknowledgments}
We thank Gustavo Vicente for useful discussions.
P.C.M.D. and N.P.N. would like to thank CAPES grant 88882.332430/2019-01 and CNPq grant PQ-IB 309073/2017-0 of Brazil, respectively, for financial support.
\end{acknowledgments}

\appendix

\end{document}